\documentclass[preprint,aps]{revtex4}

\begin{document}
\draft
\title{A review on the decoy-state method for practical quantum key distribution}
\author{Xiang-Bin Wang\thanks{Email address: wang@qci.jst.go.jp}\\
IMAI Quantum Computation and Information Project,
ERATO, JST,
Daini Hongo White Bldg. 201, \\5-28-3, Hongo, Bunkyo,
Tokyo 133-0033, Japan}
\begin{abstract} We present a review on the historic development of the decoy state method, including the background, principles, 
methods, results and development. We also clarify 
some delicate concepts. Given an imperfect source and a very lossy channel,
the photon-number-splitting (PNS) attack can make the quantum key distribution
(QKD) in practice totally insecure. Given the result of ILM-GLLP, one knows how to distill the
secure final key if he knows the fraction of tagged bits. The purpose of decoy state method is to do a tight verification
of the the fraction of tagged bits. The main idea of decoy-state method is changing the intensities of source light and one can
verify the fraction of tagged bits of certain intensity by watching the the counting rates of pulses of different intensities.
Since the counting rates are small quantities, the effect of statistical fluctuation is very important. It has been shown that
3-state decoy-state method in practice can work even with the fluctuations and other errors.
\end{abstract}
\maketitle  
\section{introduction and backgroound}
Although many standard quantum key distribution (QKD) protocols such as BB84\cite{bene} have been proven to be unconditionally
secure\cite{maye,shor2,ekert}, this does not guarantee the security of QKD in practice, due to various types of imperfections in a practical
set-up.
 In practical QKD, the source is often imperfect. Say, it may produce multi-photon
pulses with a small probability. Normally weak coherent state is used in practical QKD. The probability of multi-photon
pulses is around $10\%$ among all non-vacuum pulses. On the other hand, the channel can be very lossy. For example, if we want to
do QKD over a distance longer than 100kms, the overall transmittance can be in the magnitude order of $10^{-3}$
or even $10^{-4}$. This opens a door for the Eavesdropper (Eve) by the so called photon-number-splitting (PNS) attack\cite{bra}.

It was then shown by Inamori, L\"utkenhaus and Mayers (ILM)\cite{inl} and by Gottesman {\em et al} (GLLP)\cite{gllp} 
on how to distill the secure final key even with an imperfect source, provided that we have a way to verify the upper bound of
the fraction of tagged bits (counts caused by multiphoton pulses from the source) or equivalently, the lower bound of
untagged bits (counts caused by single-photon pulses from the source). However, the ILM-GLLP\cite{inl,gllp} result does not tell
us how to make the verification itself. What it has presented is how to make the finall key $given$ the verified results
of fraction of tagged bits or fraction of untagged bits. 
Therefore, the only difficulty remained for secure QKD in practice is the 
{\em verification}. Non-trivial verification is the central issue of the 
decoy-state method.

Before going into the decoy-state method, let's first recall some concepts 
and results of ILM-GLLP\cite{inl,gllp}.
\subsection{ Tagged bits} Suppose in a QKD protocol, 
Alice is the sender and Bob is the receiver.
In the standard protocols such as BB84 with perfect single photon source, there is no $tagged$ bits because
an Eavesdropper (Eve) in principle will cause errors to Bob's bits
if she wants to know some of the bit values of Bob. However, if Alice uses
an imperfect source, things will be different. Suppose sometimes Alice sends
a multi-photon pulse. All the photons in the pulse have the same state. Eve
can  keep one photon from the pulse and sends other photons
of the pulse to Bob. This action will not cause errors to the bit value 
to bob but Eve may have full information about Bob's bit: 
After the measurement basis is announced by Alice or Bob, Eve will be always able to measure the 
photon she has kept in the correct basis. {\em If Eve can know some of bit values without causing any errors, 
these bits are defined as tagged bits.}
Given an imperfect source, whenever Alice sends out a multi-photon pulse and Bob's detector counts, 
we assume that a tagged bit has been
produced. {We don't care how many photons the pulse may contain after it is transmitted to Bob's side.}
\\{\em Remark.} Bob cannot verify the tagged bits at his side by measuring the photon number in each comming pulses. Say, suppose
he finds certain pulse contains only one photon. The bit caused by that pulse could be still a tagged bit because the pulse could have contained
two photons at Alice's side.\\  
\subsection{Final key distillation with a fraction of tagged bits.}
In the standard BB84 protocol with perfect single photon source, there is no 
tagged bits. To distill the final key, we need the information of bit-flip 
error rate $t_b$ and {\em phase-flip} rate $t_p$. Based on this information we can
in principle have a CSS\cite{shor2} code to correct all bit-flip errors (error correction) 
and to compress any third party's information to almost zero (privacy amplication). 
In pararticular, we shall use a CSS code that consumes $n_rH(t_b)=n_r[-t_blog_2 t_b -(1-t_b)log_2(1-t_b)]$ raw bits to correct 
bit-flip errors and
$n_rH(t_p)$ raw bits for the privacy amplification. Here $n_r$ is the number of raw bits. 
It was shown by ILM-GLLP\cite{inl,gllp} that we can also distill the secure final key by a CSS code even with an imperfect source,
if we know the bit-flip rate, phase-flip rate and upper bound value of the fraction of tagged bits, $\Delta$.
In particular we shall use a CSS code that consumes
 $n_tH(t_b)$ raw bits for error correction and $n_r[\Delta +(1-\Delta)H(\frac{t_p}{1-\delta})$ for privacy amplification.
This is to say, in a protocol with perfect single-photon source, we shall know how to distill the final key if we know 
the bit-flip rate and phase-flip rate; in a protocol with an imperfect source, we shall also know how to distill the final key
if we know  the bit-flip rate, phase-flip rate and $\Delta$ value, the upper bound of the tagged bits. Equivalently, we can also
use $\Delta_1$, the lower bound of untagged bits.  It is easy to verify the bit-flip
rate and phase-flip rate: Alice and Bob simply take some samples and announce the bit values. Asymptotically, the  error rate rate
of the remained raw bits is equal to the error rate of those samples. However, to know a tight bound value of $\Delta$ or $\Delta_1$
is not so straightforward. The central task for the decoy-state method is to make a tight verification of $\Delta$ or $\Delta_1$.  
 
\section{Hwang's idea and  protocol} 
The first idea of decoy-state method and the first protocol
   is given by Hwang\cite{hwang}. Hwang proposed to do the non-trivial verification by changing the intensity of pulses.
In Hwang's first protocol, two intensities are used. The intensity
of signal pulses is set to be around 0.3 and
the intensity of decoy-state pulses is set to be 1. By watching the counting rate of decoy pulses,
one can deduce the upper bound of the fraction of tagged bits among all those signal pulses.

For clarity, we first demonstrate why a tight verification is non-trivial.
Knowing the photon-number distribution of the source and the channel transmittance to one coherent state
is insufficient for a tight verification.
Consider a coherent state with intensity $\mu$. With the phase totally randomized, the state is a probabilistic mixture of
different photon numbers:
\begin{eqnarray}
\rho_\mu=e^{-\mu}\sum_{n=0}^{\infty} \frac{\mu^n}{n!}|n\rangle\langle n| \label{source0}
\end{eqnarray}
and $|n\rangle\langle n| $ is the Fock state of $n$ photons. Since Alice does not know the photon number in each individual pulse,
their knowledge about the channel transmittance is the $ averaged$ transmittance to the whole mixed state. However, in principle,
the channel (Eve's channel) transmittance can be selective: it can be more transparent given a multi-photon pulse.
Suppose the channel transmittance is $\eta$. (Since Eve may also control Bob's detector, here $\eta$ is the overall transmittance
for channel and detector.) It is possible that the transmittance for single-photon pulses is actually zero. If we use the worst-case estimation,
we require $\mu <\eta$ in order to obtain a meaningful result for the verification.
In practice, if the distance is longer than 100kms, it is possible that $\eta$ value is in the magnitude order of $10^{-3}$ or $10^{-4}$,
then the worst-case verification doesn't work.    

Earlier, the PNS attack has been investigated where Alice and Bob 
monitor only how many non-vacuum signals arise, and how many errors 
happen. However, it was then shown\cite{kens1} that the simple-minded method
does not guarantee the final security. It is shown\cite{kens1}  that in a 
typical parameter regime nothing changes if one starts to monitor the 
photon number statistics as Eve can adapt her strategy to reshape the 
photon number distribution such that it becomes Poissonian again.
\\{\em Remark:} Here PNS attack allows any method for Eve to split the multi-photon pulses, provided that it does not 
violate laws of the nature. Although  some types of specific PNS attack, e.g., the beam-splitter attack 
could be detected by simple method such as tomography at Bob's side, a method to manage {\em whatever} type of PNS attack is
strongly non-trivial. 

However, as it was first shown by Hwang\cite{hwang}, by changing the intensity of pulses, one can make the verification 
unconditionally and more efficiently. We now start from the classical statistical principle.
\\{\em Principle 1.} Given a large number of independent pulses, the averaged value of any 
physical quantity per pulse in a randomly chosen subset of pulses must be (almost) equal to that of the remained
pulses, if the number of pulses in the subset and the number of remained pulses is large.\\
In a standard QKD protocols with perfect single-photon source, this principle is used for the error test:
They check the error rate of a random subset, and use this as the error rate of the remained bits. Also, this principle
can be used for estimation of other quantities, such as the {\em counting rate}.
 In the protocol, Alice sends pulses to Bob. Given a loss channel, whenever a pulse is sent out from Alice,
Bob's detector may click may not click. If his detector clicks, a raw bit is generated. Counting rate is the ratio 
of the number
of Bob's click and the number of pulses sent out from Alice. 
More specifically, if source $x$ sends out $N$ pulses and
Bob's detector clicks $n_x$ times meanwhile, the counting rate for pulses from source $x$ is $S_x=\frac{n_x}{N}$.\\
In the QKD protocol, Alice controls the source. Consider a case of a mixed source of X and source Y. When we say
{\em a mixed source of X and Y}, we mean that each maybe from X or Y randomly. 
If these two sources produce the {\em same} state and each pulses are independent and the number pulses from each source
is sufficiently large, then the counting rate of source X must be equal
to that of source Y, since  X and Y can be regarded as one source and pulses from X can be regarded as samples for testing
and pulses from source Y can be regarded as the remained pulses.
\\{\em Principle 2.} Asymptotically, given a mixed source of X and Y,
Alice can verify the counting rate of source Y by watching counting rate of source X, if X and Y produce the same
states and each pulses are independent.    
   
Now we see how  Hwang's original protocol\cite{hwang} works. For mathematical simplicity, we give up Hwang's original
derivation which involves counting rates of each Fock states and complicated inequalities. We use the technique of
density matrix convex and there are only a few parameters involved\cite{wang0}.
We omit the dark count at the moment and assume that a vacuum pulse from Alice never causes counts at Bob's side.
Consider the source state in equation(\ref{source0}). The state can be re-written in the following equivalent convex form:
\begin{eqnarray}
\rho_{\mu}= e^{-\mu}|0\rangle\langle0|+\mu e^{-\mu}|1\rangle\langle 1|
+c\rho_c\label{oo}
\end{eqnarray}
and $c=1-e^{-\mu}-\mu e^{-\mu}>0$, 
\begin{eqnarray}
\rho_c=\frac{1}{c}\sum_{n=2}^\infty P_n(\mu)|n\rangle\langle n|\label{coherent}
\end{eqnarray} 
and $P_n(\mu)=\frac{e^{-\mu} \mu^{-n}}{n!}$.
This convex form shows that the source sends out 3 types of pulses: sometimes sends out vacuum, sometimes sends out $|1\rangle\langle 1|$, sometimes sends
pulses of state $\rho_c$. 
Bob's counts caused by $\rho_c$ from Alice are regarded as $tagged$ bits.
Since we know explicitly the probability of pulses  $\rho_c$ for our source, we shall know the fraction of tagged bits
if we know $s_c$, the {\em counting rate} of state $\rho_c$. The {\em counting rate} of any state $\rho$ is the probability
that Bob's detector counts whenever Alice sends $\rho$. If we have another source  $A'$ which always produces state
$\rho_c$, then we can combine source A, the coherent source $\rho_\mu$ and $A'$. Say, Alice uses a mixed source of A and A'. 
After Alice sends out all pulses, Bob announces which time is counted and which time is not. Then Alice knows
the counting rate of source A'. The counting rate of source A' is just the counting rate of all pulses $\rho_c$ from source A, asymptotically. 
Since Eve cannot treat the pulses from source A' and the pulses of state $\rho_c$ from source A differently.
This can be understood more easily in the following way: the above mixed source of A and A' can be equivalently regarded as 4 sources since
source A can be equivalently regarded as 3 sub-sources: sub-source $A_0$ containing all vacuum pulses from A, sub-source $A_1$ contains all 
single-photon pulses from A and sub-source $A_c$ contains all pulses of state $\rho_c$. Since the states of pulses from source A' and source $A_c$ are 
identical, Eve can not treat them differently. Therefore the counting rate for pulses from source A' must be equal to the that of
source $A_c$. Since source $ A_c$ contains all multi-photon pulses for source A, therefore the value $s_c$ for source A is verified
by watching the counting rate of source A'. This is a natural consequence of {\em Principle 2:} Source A' and sub-source $A_c$ makes a
{\em mixed source}, they each produce the same identical pulses, therefore  
Alice can verify the counting rate of sub-source $A_c$ by watching that of source A'.

In the above toy model, we have used source A' that produces state $\rho_c$ deterministically. In practice we don't have such a source.
But we have another coherent source with a different intensity $\mu'>\mu$. We now consider a realistic mixed source: source A is the coherent
source with intensity $\mu$, source $A_{\mu'}$ is another coherent source with intensity $\mu'$. Since $\mu'>\mu$, the state for source
$A_{\mu'}$ can be written in the convex form:
\begin{eqnarray}
\rho_{\mu'}=e^{-\mu'}|0\rangle\langle0|+\mu' e^{-\mu'}|1\rangle\langle 1|
+c\frac{\mu'^2 e^{-\mu'}}{\mu^2 e^{-\mu}}\rho_c + d\rho_d\label{dd}.
\end{eqnarray} 
The source $A_{\mu'}$ can be equivalently regarded as a mixed source which contains the 4 sub-sources:
$A_{\mu'0}$ which contains all vacuum pulses from $A_{\mu'}$, $A_{\mu'1}$ which contains all single-photon pulses from $A_{\mu'}$, $A_{\mu' c}$
that contains all $\rho_c$ pulses of  source $A_{\mu'}$ and $A_{\mu' d}$ that contains all $\rho_d$ pulses of source $A_{\mu'}$.
Therefore the mixed source of $A,A_{\mu'}$ now contains 7 sub-sources.
Since state from sub-source $A_{\mu' c}$ is identical to that of sub-source $A_c$, the counting rate of source $A_{\mu' c}$ should be equal
to that of $A_c$, i.e., the counting rate of state $\rho_c$ from source A. If Alice knew which pulses were from sub-source $A_{\mu' c}$, she
would know the value $s_c$ exactly therefore know the value of fraction of tagged bits explicitly for source A, and then did key 
distillation based on raw bits from source A. However, Alice had no way to know which pulses are from sub-source $A_{\mu' c}$. But she know
which pulses are from source $A_{\mu'}$. We shall show that she can know an {\em upper bound} of the fraction of tagged bits from source A
by watching the counting rate of source  $ A_{\mu'}$.

In the protocol, Alice can watch the  (averaged) counting rate for source $A_{\mu'}$ and we soppose the value is $S_{\mu'}$. According to equation (\ref{dd}), 
we have the following equation:
\begin{eqnarray} 
S_{\mu'}= \mu' e^{-\mu'}s_1 + c\frac{\mu'^2 e^{-\mu'}}
{\mu^2 e^{-\mu}}s_c +ds_d.\label{origin}
  \end{eqnarray}  
Here we have assumed no vacuum dark count and denoted $s_d$ for the counting rate for state $\rho_d$, i.e. for source $A_{\mu' d}$
, $s_1$ for counting rate of single-photon pulses, i.e., for source $A_{\mu'}$.
Alice does not know the value of $s_1$ or $s_d$ but she knows the fact $s_1\ge 0$ and $s_d\ge 0$.
Therefore we transform eq(\ref{origin}) into an inequality for the upper bound of $s_c$:
\begin{eqnarray}
s_c
\le \frac{c\mu^2e^{-\mu}}{\mu'^2e^{-\mu'}}S_{\mu'}\label{crude}
\end{eqnarray}
This bound is obtained  based on the observation of source $A$. This is the bound value for counting rate of sub-source $A_{\mu' c}$. This
is {\em also} the bound value for state $\rho_c$ from any source, including those $\rho_c$ pulses from source A, since Eve cannot treat pulses
of the same
state differently according to which source the pulse is  from.
Therefore we have the following upper bound for fraction of tagged bits of source A:
\begin{eqnarray}
\Delta \le \frac{\mu^2e^{-\mu}S_{\mu'}}{\mu'^2e^{-\mu'}S_{\mu}},
\end{eqnarray}
and we have used
\begin{eqnarray}
\Delta=c\frac{s_c}{S_{\mu}}\label{new}
\end{eqnarray}
In the normal case that there is no
Eve's attack, Alice and Bob will find
$
S_{\mu'}/S_{\mu}=\frac{1-e^{-\eta\mu' }}{1-e^{-\eta\mu}}= \mu'/\mu
$ in their protocol therefore they can verify 
$
 \Delta \le \frac{\mu e^{-\mu}}{\mu' e^{-\mu'}}
,$
which is just eq.(13) of Hwang's work\cite{hwang}.

The above is the main result of Hwang's work. We have simplified the original derivation given by Hwang\cite{hwang}.
In summary, Hwang's protocol works in this way: Use the intensity $\mu'=1$ for the decoy state. By watching the counting rate of 
decoy state, we can obtain $\Delta$ value for the signal state (intensity $\mu$). 

Although Hwang's result of verification has been much better than the trivial 
worst-case method, Hwang's protocol should be further
improved for immediate use in practice. Hwang's verified result is still much larger than the true value. We want a {\em tighter}
estimation.
\\{\em Remark: } The security of Hwang's method is a direct consequence the separate prior art result of ILM-GLLP\cite{inl,gllp}.
ILM-GLLP\cite{inl,gllp} have offered methods to distill the unconditionally secure final key from raw key if the upper bound
of fraction of tagged bits is known, given whatever imperfect source and channel. Decoy-state method verifies such an upper bound
for coherent-state source. We can consider an analog using the model of pure water distillation: Our task is to distill pure water
by heating from raw water that may contain certain poison constitute. Surppose it is known that the poison constitute will be evaporated
in the heating. We want to know how long the heating is needed to obtain
the pure water for certain. If we blindly heat the raw water for too long, all raw water will be evaporated and we obtain nothing.
If we heat the raw water for a too short period, the water could be still poisonous. 
 ``ILM-GLLP'' finds an explicit formula for the heating time which is a function of the upper bound of the fraction
of poison constitute. They have proven that we (almost) always obtain pure water if we use that formula for the heating time.
However, the formula itself does not tell how to examine the fraction of poison constitute. ``Decoy-state'' method is a method to verify a 
tight
upper bound of the fraction of the poison constitute. It is guaranteed by the classical statistical principle that the
verified upper bound by ``decoy-state method'' is (always) larger than the true value. Using this analog, the next question is how to
obtain a tighter upper bound: if the verified value over estimates too much, it is secure but it is inefficient. We want a way to
obtain a value that is only a bit larger than the true value in the normal case that there is no Eve (for efficiency), 
{\em and} it is (almost) {\em always}  larger than
the true in whatever case (for security). This can be achieved by the improved decoy-state method.
Here the term ``(almost) always'' means ``with a probability exponentially close to 1''.   
\section{improved decoy-state method}
The improvement is possible because Hwang's method has not sufficiently using the different intensities. Actually, in doing the verification,
Alice has only used the counting rate of one intensity, the source $A_{\mu' }$. It should be interesting to consider the case 
that Alice uses more intensities. 

After Hwang's work, decoy-state method is further studied. Ref\cite{lo4} reviewed the PNS attack and 
the elementary idea of decoy-state method
with some shortly-stated suggestions for possible improvement, but there is no conclusive result. 
It suggests doing the verification by using two intensities, vacuum and very weak coherent
state. However this idea seems to be inefficient in practice due to the possible fluctuation
of dark count\cite{wangc}. Latter, a protocol with infinite number of intensities is proposed\cite{tot} and the result
in the infinity limit is given. The main result there\cite{tot} has been published in Ref\cite{lolo}.

Here we are most interested in a protocol that is {\em practically} efficient. Obviously, there should be several criterion. 1. The protocol must
be clearly stated. For example, there should be {\em quantitative} 
description about the intensities used and 
{\em quantitative} result about the verification.
Because we need the $explicit$ information of intensities in the implementation and the $explicit$ 
value of $\Delta$ for key distillation. 2. The result of verified value $\Delta$ should be tight in the normal case when there is no
Eve. This criterion is to guarantee a good final key rate. 3. It should only use a few different intensities.
4. It should be robust to possible statistical fluctuations. Say, in the non-asymptotic case, it only needs a reasonable number
of pulses to make the verification. Note that the  counting rates are very small parameters. The effects of possible
statistical fluctuations can be very important.
Concerning the above criterion, a 3-intensity protocol is then proposed\cite{wang0}. The protocol uses 3 intensities: vacuum, $\mu$ and $\mu'$
for the verification.
For convenience, we shall always assume
\begin{eqnarray}
\mu'>\mu; \mu' e^{-\mu'} > \mu e^{-\mu} \label{condition}
\end{eqnarray} in this paper. 
Since we randomly change the intensities among 3 values, we can regard it as the mixing of 3 sources.
Source $A_0$ contains those vacuum pulses, $A$ contains those pulses of intensity $\mu$ and source $A_{\mu'}$ contains those pulses of intensity 
$\mu'$. States from source $A$ and $A_{\mu'}$ are given by eq.(\ref{oo}) and eq.(\ref{dd}), respectively.
In the protocol, they can direct watch the counts of each source of $A_0,A,A_{\mu'}$. Suppose they find $S_0,S_\mu,S_{\mu'}$
for each of them. In the asymptotic case, we have the following equations:
\begin{eqnarray}
S_{\mu}= e^{-\mu}S_0 + \mu e^{-\mu}s + cs_c \label{originmu}
  \end{eqnarray} 
\begin{eqnarray}
S_{\mu'}= e^{-\mu'}s_0 + \mu' e^{-\mu'}s_1 + c\frac{\mu'^2 e^{-\mu'}}
{\mu^2 e^{-\mu}}s_c +ds_d \label{originp}
  \end{eqnarray}
In the above we have used the same notations $S_0,s_1,s_c$ in both equations. This is because we have
assumed that the counting rates of the same state from different sources are equal.
$S_0$ is known, $s_1$ and $s_d$ are unknown, but they are never less than 0. Therefore
setting $s_d, s_1$ to be zero we can obtain the following crude result by using eq.(\ref{originp}) alone.
\begin{eqnarray}
cs_c\le \frac{\mu^2e^{-\mu}}{\mu'^2e^{-\mu'}}\left(S_{\mu'}- e^{-\mu'}s_0-\mu' e^{-\mu'}s_1\right).
\label{origin8} 
\end{eqnarray}
However, we can tighten the verification by using eq.(\ref{originmu}).
Having obtained the crude results above,
 we now show that the verification can be done
more sophisticatedly and one can further tighten the bound significantly.
In the inequality (\ref{crude}), we have dropped terms $s_1$ and $s_d$, since
we only have trivial knowledge about $s_1$ and $s_d$ there,
i.e., $s_1\ge 0$ and $s_d\ge 0$. Therefore, inequality(\ref{origin8}) has no advantage at that moment.
 However, after we have obtained
the crude upper bound of $s_c$, we can have a larger-than-0 lower
bound for $s_1$, provided that our crude upper bound for $\Delta$ given by eq.(\ref{crude}) is not too large.
 From eq.(\ref{oo}) we have
\begin{eqnarray}
e^{-\mu}s_0 + \mu e^{-\mu}s_1 + c s_c=S_{\mu}. \label{crude1}
\end{eqnarray}
 With the crude upper bound for $s_c$ given
by eq.(\ref{crude}), we have the non-trivial lower bound for $s_1$ now:
\begin{eqnarray}
s_1 \ge S_{\mu}-e^{-\mu}s_0 - c s_c
 > 0. \label{new1}
\end{eqnarray}
Therefore tight 
values for $s_c$ and $s_1$ can be obtained by solving the simultaneous constraints of 
equation (\ref{crude1}) and  inequality (\ref{origin8}).
We have the following final bound after solving them:
\begin{eqnarray}
\Delta \le
\frac{\mu}{\mu'-\mu}\left(\frac{\mu e^{-\mu} S_{\mu'}}{\mu' e^{-\mu'} S_{\mu}}-1\right) 
+\frac{\mu e^{-\mu}s_0 }{ \mu' S_\mu }.\label{assym}
\end{eqnarray}
Here we have used eq.(\ref{new}). In the case of $s_0<<\eta$, 
if there is no Eve., $S_\mu'/S_\mu=\mu'/\mu$. 
Alice and Bob must be able to verify 
\begin{eqnarray}
\Delta = \left. \frac{\mu \left(e^{\mu'-\mu}-1\right)}{\mu'-\mu}\right|_{\mu'-\mu\rightarrow 0}=\mu
\end{eqnarray} 
in the protocol.
This is close to the real value of fraction of multi-photon counts: $1-e^{-\mu}$, given that $\eta<<1$. 
In this 3-intensity protocol for the verification, both $\mu$ and $\mu'$ can be set in a reasonable range
therefore both of them can be used for final key distillation. Of course, if we also want to use source
$A_{\mu'}$ for key distillation, we need the value $\Delta'$, the upper bound of the fraction
of tagged bits for source $A_{\mu'}$. Given $s_c$, we can calculate the lower bound of $s_1$ through eq.(\ref{new1}).
Given $s_1$, we can also calculate the upper bound of $\Delta'$, the fraction
of multi-photon count among all counts caused by pulses from
source ${A_{\mu'}}$. Explicitly,
\begin{eqnarray}
\Delta' \le 1- (
1-\Delta -\frac{e^{-\mu}s_0}{S_\mu})e^{\mu-\mu'}-\frac{e^{-\mu'}s_0}{S_{\mu'}}.
\end{eqnarray}
The values of  $\mu,\mu'$ should be chosen in a reasonable range, e.g., from 0.2 to 0.5.
\section{statistical fluctuations}
The results above are only for the asymptotic case. 
In practice, there are statistical fluctuations, i.e., Eve. has non-negligibly
small probability to
treat the pulses from different sources a little bit differently,
even though the pulses have the same state.
Mathematically, this can be stated by
\begin{eqnarray}
s_{\rho}(\mu')=(1+r_\rho)s_\rho(\mu)
\end{eqnarray} 
and the real number $r_\rho$ is the relative statistical fluctuation
for counting rate of stte $\rho$ in different sources of pulses.
It is $insecure$ if we simply use the asymptotic result in practice.
Since the actual values are actually different from what we have estimated
from the observed data.
Our task remained is  to verify a tight upper bound
of $\Delta$ and the probability that the real value of $\Delta$ breaks the verified upper bound
is exponentially close to 0. 

The counting rate of any state $\rho$ from different sources
now can be slightly different from the counting rate of the same state $\rho$ 
from another sources, $A_\mu$, with non-negligible
probability. We shall use the primed
notation for the counting rate for any state from source $A_{\mu'}$ and the original notation
for the counting rate for any state from source $A$. 
Explicitly, constraints (\ref{crude},\ref{new1})
are now converted to
\begin{eqnarray}
  \left\{ \begin{array}{l} 
e^{-\mu}s_0 + \mu e^{-\mu}s_1 + c s_c=S_{\mu},
 \\ cs'_{c}\le \frac{\mu^2e^{-\mu}}{\mu'^2e^{-\mu'}}\left(S_{\mu'}
- \mu' e^{-\mu'} s'_1
- e^{-\mu'}s'_0\right) .
  \end{array}
  \right. \label{couple}
 \end{eqnarray}
Setting $s_x' = (1-r_x)s_x$ for $x=1,c$  and $s'_0=(1+r_0)s_0$ with $r_x> 0$ we obtain
\begin{eqnarray}
\mu' e^\mu \left[(1-r_c)\frac{\mu'}{\mu}-1\right]\Delta \le \mu e^{\mu'}S_{\mu'}/S_\mu
-\mu'e^{\mu}+[(\mu'-\mu)s_0+r_1s_1+r_0s_0]/S_\mu.
\end{eqnarray}
From this we can see, if $\mu$ and $\mu'$ are too close, $\Delta$ can be very large. 
The important question here is now whether there are reasonable
values for $\mu',\mu$ so that our method has significant advantage to the
previous method\cite{hwang}. The answer is yes.

Given $N_1+N_2$ copies of state $\rho$,  suppose
the counting rate
for $N_1$ randomly chosen states is $s_{\rho}$ and the counting rate
for the remained states  is $s'_{\rho}$, the probability that $s_\rho-s'_\rho>\delta_\rho$
is
less than $\exp\left(-\frac{1}{4}{\delta_\rho}^2N_0/s_\rho\right)$
and $N_0 ={\rm Min}(N_1,N_2)$. Now we consider the difference of counting rates
for the same state from different sores, $A$ and $A_{\mu'}$.
 To make a faithful estimation 
for exponentially sure, we require 
${\delta_\rho}^2N_0/s_\rho =100$. This causes a relative fluctuation   
\begin{eqnarray}
r_\rho=\frac{\delta_\rho}{s_{\rho}}\le 10\sqrt{\frac{1}{s_{\rho}N_0}}\label{statis}.
\end{eqnarray} 
The probability of violation is less than $e^{-25}$.
 To formulate the relative fluctuation $r_1,r_c$
by $s_c$ and  $s_1$, we only need to check the number of pulses in $\rho_c$,
$|1\rangle\langle 1|$  in each sources in the protocol.
That is, using eq.(\ref{statis}),
we can replace $r_1,r_c$ in eq.(\ref{couple}) 
$|1\rangle\langle 1|$  in each sources in the protocol.
That is, using eq.(\ref{statis}),
we can replace $r_1,r_c$ in eq.(\ref{couple}) 
by $10e^{\mu/2}\sqrt{\frac{1}{\mu s_1N}}$, 
$10\sqrt{\frac{1}{c s_cN}}$, respectively
and $N$ is the number of pulses in source $A$.
Since we assume the case where vacuum-counting rate is much less than
the counting rate of state $\rho_\mu$, we omit the effect of fluctuation
in vacuum counting, i.e., we set $r_0=0$.
 With these inputs, eq.(\ref{couple}) can now be solved
numerically.
 The results are listed in the following table. 
 From this table we can
see that good values of $\mu,\mu'$ indeed exist and our verified
upper bounds are sufficiently tight to make QKD over very lossy channel.
Note that so far this is the $only$ non-asymptotic result among all existing works on decoy-state. 
From the table we can see that our non-asymptotic values are less than Hwang$'$s asymptotic
values already. Our verified values are rather close to the true values. We have assumed the vacuum count rate
$s_0=10^{-6}$ in the calculation. If $s_0$ is smaller, our results will be even better. 
Actually, the value of $s_0$ (dark count) can be even
lower than the assumed value here\cite{gobby,tomita}. 
\begin{table}
\caption{The verified upper bound of
the fraction of tagged pulses in QKD.
$\Delta_H$ is the result from Hwang's method. $\Delta_R$ is the true value of the fraction of
multi-photon counts in case there is no Eve.
 $\Delta_H$ and $\Delta_R$ do not change with channel transmittance.
$\Delta_{W1}$ is bound for pulses
from source $A$, given that $\eta=10^{-3}$.
 $\Delta_{W2}$ and $\Delta'_{W2}$  are bound values
for the pulses from source $A,A_{\mu'}$ respectively,
given that $\eta=10^{-4}$. 
We assume $s_0=10^{-6}$. The number of pulses is
 $ 10^{10}$  from source $A,A_\mu'$ in calculating $\Delta_{W1}$
and $8\times 10^{10}$ in calculating $\Delta_{W2},\Delta'_{W2}$.
(Our results will only increase by 0.03 even if we only use $10^{10}$ pulses. Actually, as we shall show it in Table 2 and 3, 
pretty good results
can be obtained with only $10^{10}$ pulses\cite{wang2}.) 
$4\times 10^{9}$ 
vacuum pulses
is sufficient for source $A_0$. The bound values will change by less than
0.01 if the value of $s_0$ is 1.5 times larger. The numbers inside
brackets are chosen values for $\mu'$. 
For example, in the column of $\mu=0.25$,
data $30.9\%(0.41)$ means, if we choose $\mu=0.25,\mu'=0.41$, we can
verify $\Delta\le 30.9\%$ for source $A$. }
\begin{tabular}{c|c|c|c|c}
$\mu$ & 0.2 & 0.25 &0.3 & 0.35  
\\ 
\hline 
$\Delta_H$ & 44.5\% &52.9\%&  60.4\% & 67.0\%\\ \hline
$\Delta_{R}$ & 18.3\% & 22.2\% & 25.9\%
& 29.5\%\\
\hline
$\Delta_{W1}$ & 23.4\%(0.34)& 28.9\%(0.38)& 34.4\%(0.43)
& 39.9\%(0.45)\\
\hline
$\Delta_{W2}$ & 25.6\%(0.39) & 30.9\%(0.41) & 36.2\%(0.45
)& 41.5\%(0.47)
\end{tabular}
\begin{tabular}{c|c|c|c|c}
$\mu'$ & 0.39 & 0.41 &0.45 & 0.47\\ \hline
$\Delta_H$ & 71.8\% & 74.0\% & 78.0\% & 79.8\%
\\ \hline
$\Delta_{R}$ & 32.3\% & 33.7\% & 36.2\%
& 37.5\% \\ \hline
$\Delta'_{W2}$ & 40.1\% & 42.2\% & 45.8\% & 48.6
\end{tabular}
\end{table}
In the real set-up given by Gobby et al\cite{gobby}, the light losses a half
over every 15km, the devices and detection loss is $4.5\%$ and
$s_0\le 8.5\times 10^{-7}$. 
Given these parameters, we believe that our protocol
works over a distance longer than 120km with 
 with $\mu=0.3,\mu'=0.45$ and {\em a reasonable number} of total pulses. 
In the table, we have chosen both values of $\mu,\mu'$ in a reasonable range. Of course, our method and
Eq.(\ref{assym}) also work for other values of $\mu,\mu'$ which are beyond the table. 
This shows that eq.(\ref{assym}) indeed gives a rather tight upper bound. 
\section{Robustness to other small errors}
We now study how robust our method is.
In the protocol, we use different intensities. 
In practice, there are both statistical fluctuations and small operational errors in switching
the intensity.
We shall show that, by using the counting rates of 3 intensities, one can still verify 
tight bounds even we take all theses errors and fluctuations into consideration.

There are small operational errors inevitably. Say, in setting the intensity of any light pulse,
the actual intensity can be slightly different from the one we have assumed.
More specifically, suppose the number of pulses from source $A_0,A,A_{\mu'}$ are
$N_0,N_\mu,N_{\mu'}$, respectively.
 
Due to the small operational error, the intensity of light pulses in source $A_0$ could be
slightly larger than 0. This doesn't matter because a little bit over estimation on the vacuum count
will only decrease the efficiency a little bit but not at all undermine the security. Therefore
we don't care about the operational error of this part. Say, given $n_0$ counts for all the pulses
from source $A_0$, we then simply assume the tested vacuum counting rate is
$s_0=n_0/N_0$, though we know that the actual value of vacuum counting rate is less than this.

We shall only consider sources $A,A_{\mu'}$. 
First, there are statistical fluctuations to the states itself since the number of pulses are finite.
Say, e.g., given $N_\mu$ pulses of intensity $\mu$, the number of vacuum, single-photon state and multi-photon
state $\rho_c$ could be a bit different from the assumed values of $P_0(\mu), P_1(\mu),P_c(\mu)$, respectively.
The similarly deviations also apply to the pulses of intensity $\mu'$. But the deviation should be small given
that the number of pulses in each sources is not too small. Say, e.g., given $N_\mu\ge  10^9,\mu=0.2$,
the relative fluctuation for the probability of state $\rho_c$ is less than
$10\sqrt\frac{1}{N_\mu (1-e^{-\mu}-\mu e^{-\mu})}<0.2\%$.
Besides this, there are operational errors.   
At any time Alice decides to set the intensity of the pulse
to be $\mu$ or $\mu'$, the actual intensity could be $\mu_i$ or $\mu_i'$, which can be a bit different
from $\mu$ or $\mu'$. 
Therefore we should replace $P_\mu(n)$ and $P_{\mu'}(n)$ by slightly different distributions
of $\tilde P_\mu(n)=\frac{1}{N_\mu}\sum_{i=0}^{N_\mu}P_{\mu_i}(n) = P_\mu(n)(1+\epsilon_n)$ and 
$\tilde P_{\mu'}(n)\frac{1}{N_{\mu'}}\sum_{i=0}^{N_{mu'}}P_{\mu_i'}(n)=P_{\mu'}(n)(1+\epsilon_n')$.
 Base on these, we have the following new convex formula for the actual states of each source  given
both statistical errors of states and operational errors:
\begin{eqnarray}
\tilde \rho_\mu = \tilde P_0 |0\rangle\langle 0| +\tilde P_1 |1\rangle\langle 1| +\tilde c \tilde \rho_c
\end{eqnarray}
and $\tilde P_{0,1}=\tilde P_\mu(0,1)$; $\tilde c= 1- \tilde P_0-\tilde P_1$.
Since the new distributions are only a bit different from the old ones, there must exist
a positive number $\tilde d$ and a density operator $\tilde \rho_d$ so that the state from source
$A_{\mu'}$ is convexed by
\begin{eqnarray}
\tilde \rho_{\mu'} 
= \tilde P_0' |0\rangle\langle 0| +\tilde P_1' |1\rangle\langle 1| +\tilde {c'} \tilde \rho_c +\tilde d \tilde \rho_d.
\end{eqnarray}
This, together with the possible statistical fluctuation to counting rates gives the following simultaneous constraint:
\begin{eqnarray}
 \left\{ \begin{array}{l} 
P_\mu(0)(1+\epsilon_0)s_0 + P_\mu(1)(1+\epsilon_1)s_1 + c(1+\epsilon_c) s_c=S_{\mu},
 \\P_{\mu'}(0)(1+\epsilon_0')s'_0+P_{\mu'}(1) (1+\epsilon_1') s'_1
+ \frac{\mu'^2e^{-\mu'}}{\mu^2e^{-\mu}}c (1+\epsilon_c')s'_{c}\le S_{\mu'}
  \end{array}
  \right. \label{couple1}
\end{eqnarray}
and $c=1-P_\mu(0)-P_\mu(1);c'=\frac{\mu'^2e^{-\mu'}}{\mu^2e^{-\mu}}c$.
Suppose we know the upper bounds of all values of  $|\epsilon_x|,|\epsilon'_x|$,
$x=0,1,c$, we can calculate the lower bound of $s_1$ and upper bound of $s_c$. Say, we try all possible values
of $\epsilon_x,\epsilon'_x$ and take the worst case as the verified result.
After calculation, we find the result does not change too much given small  $|\epsilon_x|,|\epsilon'_x|$. For example, given that
 $|\epsilon_x|<2\%,|\epsilon'_x|<2\%$ and $N_\mu=N_{\mu'}=10^{10}$, $N_0=4\times 10^{9}$ and $S_0=10^{-6}$, the lower bound of
single-photon counting rate can be verified to be larger than
$0.95 \tilde s_1$ and $\tilde s_1$ is the lower bound of single-photon counts given $\epsilon_x=\epsilon'=0$. 
\section{Further studies}
After the major works presented in\cite{hwang,wang0,lolo}, the decoy-state method has been further studied. 
Harrington studied the effect of fluctuation of the state itself. Ref.\cite{wang2} 
proposed a 4-state protocol: using 3 of them to make optimized
verification and using the other one $\mu_s$ as the main signal pulses. This is because, if we want to optimize the verification
of $\Delta$ value, $\mu,\mu'$ cannot be chosen freely. Therefore we use another intensity $\mu_s$ to optimize the final key rate. 
It is shown numerically on how to choose the intensity for the main signal pulses ($\mu_s$) and good key rates are obtained in a number
of specific conditions.
Ref.\cite{05} further studied the 3-intensity protocol. In particular, statistical fluctuations to the bit error rates are also considered 
there and a type
of stronger key rate formula is suggested there. An experiment was also done\cite{exp}.
\section{summary}
In summary, we have reviewed the historic development of the decoy state method, including the background, development and 
some delicate concepts. Given an imperfect source and a very lossy channel,
the PNS attack can make the QKD in practice totally insecure. Given the result of ILM-GLLP\cite{inl,gllp}, one knows how to distill the
secure final key if he knows the fraction of tagged bits. The purpose of decoy state method is to do a tight verification
of the the fraction of tagged bits. The main idea of decoy-state method is changing the intensities of source light and one can
verify the fraction of tagged bits of certain intensity by watching the the counting rates of pulses of different intensities.
Since the counting rates are small quantities, the effect of statistical fluctuation is very important. It has been shown that
3-state decoy-state method in practice can work even with the fluctuations and other errors.

\end{document}